\documentclass[10pt,fleqn]{article}
\usepackage[paperwidth=192mm,paperheight=262mm,left=13.5mm,right=13.5mm,
 top=19mm,bottom=18mm,headheight=12pt,headsep=7pt,footskip=18pt]{geometry}
\usepackage[T1]{fontenc}
\usepackage{amsmath,amssymb,amsfonts,mathtools,amsthm}
\usepackage{tikz-cd}
\usepackage{txfonts}
\usepackage{booktabs,array,enumitem,microtype,needspace,float,mathrsfs}
\usepackage[numbers,sort&compress]{natbib}
\usepackage{etoolbox,xcolor,xurl,fancyhdr}
\usepackage[hyperfootnotes=false,hypertexnames=false,hidelinks]{hyperref}
\hypersetup{bookmarksnumbered=true,bookmarksdepth=2,
 pdfauthor={Yangcheng Li and Pingzhi Yuan},
 pdftitle={Cyclically covering subspaces and additive perfect codes},
 pdfsubject={Version 32: cyclic coverings and arithmetic criteria for perfect codes}}
\numberwithin{equation}{section}
\newcommand{\BibliographyHref}[2]{\OriginalHref{#1}{\nolinkurl{#2}}}
\AtBeginEnvironment{thebibliography}{\urlstyle{same}\let\OriginalHref\href\let\href\BibliographyHref}
\pretocmd{\section}{\Needspace{12\baselineskip}}{}{}
\pretocmd{\subsection}{\Needspace{11\baselineskip}}{}{}
\AtBeginEnvironment{theorem}{\Needspace{6\baselineskip}}
\AtBeginEnvironment{proposition}{\Needspace{5\baselineskip}}
\AtBeginEnvironment{corollary}{\Needspace{5\baselineskip}}
\AtBeginEnvironment{lemma}{\Needspace{5\baselineskip}}
\AtBeginEnvironment{example}{\Needspace{5\baselineskip}}
\AtBeginEnvironment{question}{\Needspace{5\baselineskip}}
\newtheorem{theorem}{Theorem}[section]
\newtheorem{proposition}[theorem]{Proposition}
\newtheorem{lemma}[theorem]{Lemma}
\newtheorem{corollary}[theorem]{Corollary}
\theoremstyle{definition}
\newtheorem{definition}[theorem]{Definition}

\newtheorem{example}[theorem]{Example}

\newcommand{\F}{\mathbb F}

\newcommand{\Nrm}{\operatorname{N}}
\newcommand{\wt}{\operatorname{wt}}
\newcommand{\codim}{\operatorname{codim}}
\newcommand{\ord}{\operatorname{ord}}
\newcommand{\Span}{\operatorname{span}}
\newcommand{\GL}{\operatorname{GL}}
\newcommand{\core}{\mathcal K}
\newcommand{\cA}{\mathcal A}
\newcommand{\im}{\operatorname{im}}
\newcommand{\rank}{\operatorname{rank}}

\begin{document}

\title{Cyclically covering subspaces and additive perfect codes}
\author{Yangcheng Li\thanks{Corresponding author. Email: \texttt{liyc@m.scnu.edu.cn}.}\quad
Pingzhi Yuan\thanks{Email: \texttt{yuanpz@scnu.edu.cn}.}\\[3pt]
\small School of Mathematical Sciences, South China Normal University\\
\small Guangzhou 510631, Guangdong, China}
\date{}
\maketitle
\begin{abstract}
Let $h_q(n)$ be the largest codimension of a subspace of $\F_q^n$
whose cyclic shifts cover the whole space.
For $r>0$, we characterize the inequality $h_q(n)\ge r$ in terms
of cyclic $\F_q$-linear codes over vector alphabets
with covering radius one, no nonzero word of block weight one,
and a scalar dual generated by the shifts of one word.
For $k,m\ge2$, let $I_q(j)$ count the monic irreducible polynomials
of degree $j$ over $\F_q$.
If $k$ is prime or $m\ge2\sum_{j=1}^{k-1}I_q(j)+2$, we prove
that a cyclic $\F_q$-linear perfect code of length
$\frac{q^{km}-1}{q^k-1}$ over $\F_q^k$ exists if and only if
$\gcd(m,q-1)=1$. When this condition holds, we also obtain
\(h_q\!\left(\frac{q^{km}-1}{q^k-1}\right)=k(m-1).\)
Restriction of scalars gives the same criterion for every $m\ge2$
when $q$ is odd, $q+1$ is a power of two, and $k=2p$ with $p$ prime.
We also show that $h_q(q^k+1)=k$ exactly when $q$ is even,
for every $k\ge1$.
\end{abstract}

\section{Introduction}\label{sec:introduction}

How small can a subspace of $\F_q^n$ be if its cyclic shifts
cover the whole space? Let $q$ be a prime power, $n\ge2$, and
\[
 \tau(x_0,x_1,\ldots,x_{n-1})=(x_{n-1},x_0,\ldots,x_{n-2}).
\]
A subspace $U\le\F_q^n$ is \emph{cyclically covering} if
\begin{equation}\label{eq:cover-definition}
 \F_q^n=\bigcup_{i=0}^{n-1}\tau^iU.
\end{equation}
Following Cameron, Ellis and Raynaud~\cite{CER}, let $h_q(n)$
be the largest codimension of such a subspace. Their counting
bound~\cite[Lemma~4]{CER} gives
$h_q(n)\le\lfloor\log_q n\rfloor$; we call equality
\emph{numerical extremality}.

The problem originates in Cameron's 1991 question~\cite[Problem~190]{CameronProblems}
whether $h_2(n)$ tends to infinity through odd integers, motivated
by lower bounds for Isbell's conjecture. Cameron, Ellis and
Raynaud~\cite{CER} established general bounds and constructions,
including $h_q(q^d-1)=d-1$ and
\[
 h_q\left(\frac{q^{km}-1}{q^k-1}\right)=k(m-1)
 \quad\text{if }\gcd(m,q^k-1)=1.
\]
Aaronson, Groenland and Johnston~\cite{AGJ} proved, among other
bounds, $h_q(ab)\ge h_q(a)+h_q(b)$ and $h_2(\ell)=2$ for
primes $\ell>3$ with $2$ a primitive root modulo $\ell$.
Under Artin's conjecture, this answers Cameron's question
negatively. Sun, Ma and Zeng~\cite{SMZ} obtained further
finite-field constructions and exact values, including $h_2(21)=4$.

Another line of work characterizes $h_q(n)=0$. Huang~\cite{Huang}
gave a trace criterion for $\gcd(q,n)=1$. Writing $q=p^e$,
S.~Li and Yuan~\cite{LYNoncoprime} proved that
$h_q(np^s)=0$ if and only if $h_q(n)=0$ for $\gcd(n,p)=1$.
Li, Yuan, Li and Zeng~\cite{LYLZ} obtained further vanishing
criteria and base-field relations. Wang and Shi~\cite{WangShi}
used cyclic and constacyclic codes to derive support-weight
bounds, new vanishing families, and $h_2(2\ell)=2$ under the
same primitive-root hypothesis.

We begin with a characterization of the condition \(h_q(n)=0\) in terms of full-weight words in cyclic codes of length \(n\) over \(\mathbb F_q\).
Li and Yuan gave a Fourier description in the coprime
case~\cite{LYDFT}; their cyclic-code approach~\cite{LYCC}
subsequently established, without that restriction,
\[
 h_q(n)=0
 \quad\Longleftrightarrow\quad
 \text{every nonzero cyclic code in $\F_q^n$
 contains a word of weight $n$.}
\]
This suggests the positive-codimension question:
\begin{quote}
\emph{Which coding-theoretic conditions characterize a
codimension-$r$ cyclic covering, and how can they determine $h_q(n)$?}
\end{quote}

Theorem~\ref{thm:general-bridge} characterizes positive covering
codimension through cyclic $\F_q$-linear codes over vector
alphabets, with covering radius one, no weight-one word, and a
one-generator condition on the scalar dual. The key invariant is
the \emph{cyclic core}
$\core(U)=\bigcap_{i=0}^{n-1}\tau^iU$, the largest cyclic code
contained in $U$; the quotient by this core supplies the syndrome
space of the associated code.

We next ask when the associated radius-one code is perfect,
so that every syndrome has a unique error representative of
block weight at most one. Counting gives the Hamming parameters
\begin{equation}\label{eq:intro-hamming}
	n=N_{q,k,m}=\frac{q^{km}-1}{q^k-1},\qquad r=k(m-1),
	\qquad k\ge1,\quad m\ge2.
\end{equation}
Let $I_q(j)$ denote the number of monic irreducible polynomials
of degree $j$ over $\F_q$, and put
$S_{q,k}=\sum_{j=1}^{k-1}I_q(j)$.
For $k\ge2$, Theorem~\ref{thm:prime-block} shows that, if $k$ is
prime or $m\ge2S_{q,k}+2$, a cyclic $\F_q$-linear perfect code
over $\F_q^k$ exists at this length if and only if $\gcd(m,q-1)=1$.
When this condition holds, we construct a $k$-dimensional
$\F_q$-subspace of $\F_{q^{km}}$ on which the norm to
$\F_{q^k}$ is bijective.
This also gives a codimension-$r$ cyclic covering with core
codimension $km$, and hence $h_q(n)=r$.
The criterion extends the earlier sufficient condition
$\gcd(m,q^k-1)=1$~\cite{CER}.
Corollary~\ref{cor:scalar-restriction} gives further existence
parameters by restriction of scalars. In particular, the same
criterion holds for every $m\ge2$ when $q$ is odd, $q+1$ is a
power of two, and $k=2p$ with $p$ prime.

At the boundary length $q^k+1$, corresponding to $m=2$,
Corollary~\ref{cor:boundary-application} gives, for every $k\ge1$,
\[
h_q(q^k+1)=k\quad\Longleftrightarrow\quad q\text{ is even}.
\]
Every codimension-$k$ covering at this length has core codimension
$2k$ and produces a perfect normalized code.
The existence direction is contained in~\cite[Theorem~8]{CER};
we also prove the converse and the assertion about the cyclic core.

Section~\ref{sec:structure} establishes the covering--code
correspondence and its two constructions.
Section~\ref{sec:perfect} derives the perfect-code parameters,
proves the arithmetic criterion in prime alphabet dimension and
in large extension degree, and gives the scalar-restriction and
boundary applications.
Section~\ref{sec:conclusion} summarizes the results.

\section{The covering--code correspondence}\label{sec:structure}

Throughout this paper, we consider $\F_q$-linear codes over the
vector alphabet $\cA=\F_q^a$, where $a\ge1$. Such a code is an
$\F_q$-linear subspace $P\le\cA^n$. We also call these codes
\emph{additive}, with $\F_q$-linearity understood
throughout~\cite{Huffman,KrotovMogilnykh}.
An identification $\cA\cong\F_{q^a}$ does not impose
$\F_{q^a}$-linearity. For $x,y\in\cA^n$, we use block Hamming
weight and distance,
\[
\wt(x)=|\{i:x_i\ne0\}|,\qquad d_H(x,y)=\wt(x-y),
\qquad \rho(P)=\max_{x\in\cA^n}\min_{c\in P}d_H(x,c).
\]
The right cyclic shift on $\mathcal A^n$ is
\[
\sigma(x_0,\ldots,x_{n-1})
=(x_{n-1},x_0,\ldots,x_{n-2}).
\]
We call $P$ \emph{cyclic} if $\sigma P=P$.
Here cyclicity refers to the $n$ symbols in the vector
alphabet $\mathcal A=\F_q^a$.
Positions are indexed by $\mathbb Z/n\mathbb Z$, and all dimensions
and codimensions below are over $\F_q$.

Fix an ordered $\F_q$-basis of $\cA$, using the same basis in every
block, and write $x_i=(x_{i1},\ldots,x_{ia})$ in this basis.
The \emph{scalar Euclidean dual} is
\[
P^\perp=\{y\in\cA^n:\langle x,y\rangle=0\text{ for all }x\in P\},
\qquad
\langle x,y\rangle
=\sum_{i=0}^{n-1}\sum_{j=1}^a x_{ij}y_{ij}.
\]
Its \emph{one-generator quasi-cyclic} condition means
\begin{equation}\label{eq:one-generator}
	P^\perp=\Span_{\F_q}\{z,\sigma z,\ldots,\sigma^{n-1}z\}
	\quad\text{for some }z\in P^\perp.
\end{equation}
We use $h_q(1)=0$ to include length one in the following statement.

\begin{theorem}\label{thm:general-bridge}
	Let $q$ be a prime power, $n\ge1$, and $r\ge1$. Then $h_q(n)\ge r$
	if and only if, for some $1\le a\le n-r$, there is an $\F_q$-linear
	code $P\le(\F_q^a)^n$ such that
	\begin{equation}\label{eq:radius-one-conditions}
		\sigma P=P,\qquad \rho(P)=1,\qquad \codim P=r+a,
	\end{equation}
	$P$ has no nonzero word of block weight one, and its scalar dual
	satisfies \eqref{eq:one-generator}. The code has size
	$|P|=q^{a(n-1)-r}$.
\end{theorem}

We prepare two lemmas for the proof. The first describes the
quotient determined by a cyclic core; the second records the
syndrome criteria used to pass between coverings and codes.

\begin{lemma}\label{lem:core-quotient}
	Let $n\ge1$ and $U\le V=\F_q^n$. Put
	\[
	K=\core(U)=\bigcap_{i=0}^{n-1}\tau^iU,\qquad
	E=V/K,\qquad A=U/K,
	\]
	and let $\pi:V\to E$ be the quotient map. Then the following
	statements hold.
	\begin{enumerate}[label=\textup{(\roman*)}]
		\item The space $K$ is the largest cyclic code contained in $U$.
		
		\item The shift $\tau$ induces a linear operator $T$ on $E$,
		defined by
		\[
		T(v+K)=\tau v+K.
		\]
		It satisfies
		\[
		T^n=I_E,\qquad \pi\tau=T\pi,\qquad
		\pi^{-1}(T^iA)=\tau^iU
		\quad(0\le i<n).
		\]
		If $e_0,\ldots,e_{n-1}$ is the standard basis of $V$, then
		$\pi(e_0)$ is a \emph{cyclic vector} for $T$, meaning that
		\[
		E=\Span_{\F_q}
		\{\pi(e_0),T\pi(e_0),\ldots,T^{n-1}\pi(e_0)\}.
		\]
		
		\item The shifted images of $A$ have zero common intersection,
		and they cover $E$ exactly when the shifts of $U$ cover $V$:
		\begin{equation}\label{eq:core-quotient-cover}
			\begin{aligned}
				\bigcap_{i=0}^{n-1}T^iA&=\{0\},\\
				U\text{ is cyclically covering}
				&\quad\Longleftrightarrow\quad
				E=\bigcup_{i=0}^{n-1}T^iA.
			\end{aligned}
		\end{equation}
	\end{enumerate}
\end{lemma}

\begin{proof}
	\textup{(i)}
	Since $\tau^n=I_V$, applying $\tau$ permutes the subspaces defining
	$K$, so $\tau K=K$. Also $K\le U$, because the intersection includes
	the term with $i=0$. Thus $K$ is a cyclic code contained in $U$.
	If $L\le U$ is any cyclic code, then
	$L=\tau^iL\le\tau^iU$ for every $i$. Hence $L\le K$, proving
	maximality.
	
	\textup{(ii)}
	The formula for $T$ is well defined: if $v-w\in K$, then
	$\tau v-\tau w\in K$ because $\tau K=K$.
	The identities $T^n=I_E$ and $\pi\tau=T\pi$ follow directly
	from the definition, giving the commutative diagram
	\[
	\begin{tikzcd}[row sep=large,column sep=large]
		V \arrow[r,"\tau"] \arrow[d,"\pi"'] & V \arrow[d,"\pi"]\\
		E \arrow[r,"T"'] & E.
	\end{tikzcd}
	\]
	In particular, $T$ is invertible.
	
	For each $i$, we have $\pi(\tau^iU)=T^iA$.
	If $\pi(v)\in T^iA$, choose $w\in\tau^iU$ with
	$\pi(v)=\pi(w)$. Then $v-w\in K\le\tau^iU$, so
	$v\in\tau^iU$. This proves
	$\pi^{-1}(T^iA)=\tau^iU$.
	
	Finally, $\tau^ie_0=e_i$ gives
	$\pi(e_i)=T^i\pi(e_0)$. Since $\pi$ is onto, the images of the
	standard basis span $E$, proving the cyclic-vector assertion.
	
	\textup{(iii)}
	Taking intersections of the preimage identities gives
	\[
	\pi^{-1}\left(\bigcap_{i=0}^{n-1}T^iA\right)
	=\bigcap_{i=0}^{n-1}\tau^iU
	=K=\pi^{-1}(\{0\}).
	\]
	Surjectivity of $\pi$ therefore yields
	$\bigcap_iT^iA=\{0\}$.
	Similarly,
	\[
	\pi^{-1}\left(\bigcup_{i=0}^{n-1}T^iA\right)
	=\bigcup_{i=0}^{n-1}\tau^iU.
	\]
	Since $\pi$ is onto, $\bigcup_iT^iA=E$ if and only if
	$\bigcup_i\tau^iU=V$.
\end{proof}

\begin{lemma}\label{lem:syndrome-criteria}
	Let $n,a\ge1$, let $E$ be an $\F_q$-vector space, and let
	$H:(\F_q^a)^n\to E$ be an onto linear map. Put $P=\ker H$.
	We call $H(x)$ the \emph{syndrome}
	of $x$ and refer to $E$ as the \emph{syndrome space}.
	Two words have the same syndrome if and only if their difference
	belongs to $P$.
	
	For each $0\le i<n$, let
	$\iota_i:\F_q^a\to(\F_q^a)^n$ be the block embedding:
	$\iota_i(b)$ has symbol $b$ in position $i$ and zero symbols
	elsewhere. Define
	\begin{equation}\label{eq:one-block-syndromes}
		H_i=H\circ\iota_i:\F_q^a\longrightarrow E,\qquad
		A_i=\im H_i.
	\end{equation}
	Thus $H_i(b)$ is the syndrome of the error $\iota_i(b)$, and
	$A_i$ consists of all syndromes of errors supported in position $i$.
	For $x=(x_0,\ldots,x_{n-1})$, linearity gives
	\[
	H(x)=\sum_{i=0}^{n-1}H_i(x_i).
	\]
	Then the following statements hold.
	\begin{enumerate}[label=\textup{(\roman*)}]
		\item
		\begin{equation}\label{eq:syndrome-cover}
			\begin{aligned}
				\rho(P)\le1
				&\quad\Longleftrightarrow\quad
				E=\bigcup_{i=0}^{n-1}A_i,\\
				P\text{ has no nonzero word of block weight one}
				&\quad\Longleftrightarrow\quad
				H_i\text{ is injective for every }i.
			\end{aligned}
		\end{equation}
		
		\item The code $P$ is cyclic if and only if there is a
		linear operator $T:E\to E$ satisfying $H\sigma=TH$.
		In that case, $T$ is unique and satisfies
		\begin{equation}\label{eq:induced-action}
			H\sigma=TH,\qquad T^n=I_E,\qquad
			H_i=T^iH_0,\qquad A_i=T^iA_0.
		\end{equation}
		In particular, $T$ is invertible.
		
		\item If $P$ is cyclic and $T$ is the operator in
		\textup{(ii)}, then $T$ has a cyclic vector if and only if
		$P^\perp$ satisfies \eqref{eq:one-generator}.
	\end{enumerate}
\end{lemma}

\begin{proof}
	\textup{(i)}
	Suppose first that $E=\bigcup_iA_i$. For any word
	$x\in(\F_q^a)^n$, choose $i$ and $b\in\F_q^a$ such that
	$H(x)=H_i(b)$.
	The error $e=\iota_i(b)$ has block weight at most one and
	satisfies $H(e)=H(x)$. Hence $x-e\in P$, proving
	$\rho(P)\le1$.
	
	Conversely, suppose $\rho(P)\le1$. Given $y\in E$, surjectivity
	provides a word $x$ with $H(x)=y$. Choose $c\in P$ such that
	$e=x-c$ has block weight at most one. Then
	$y=H(x)=H(e)\in\bigcup_iA_i$, proving the first equivalence.
	
	For the second equivalence, observe that
	\[
	\iota_i(b)\in P
	\quad\Longleftrightarrow\quad
	H_i(b)=0.
	\]
	Thus a nonzero word of block weight one belongs to $P$ precisely
	when some $H_i$ has a nonzero kernel.
	
	\textup{(ii)}
	Suppose $\sigma P=P$, and define
	\[
	T(H(x))=H(\sigma x).
	\]
	This is well defined: if $H(x)=H(y)$, then $x-y\in P$,
	so $\sigma(x-y)\in P$ and $H(\sigma x)=H(\sigma y)$.
	The resulting map is linear and is unique because $H$ is onto.
	Iterating the identity $H\sigma=TH$ gives
	\[
	T^nH=H\sigma^n=H.
	\]
	Surjectivity of $H$ therefore yields $T^n=I_E$, so $T$ is
	invertible.
	
	Conversely, suppose $H\sigma=TH$ for a linear operator $T$.
	If $x\in P$, then
	$H(\sigma x)=TH(x)=0$, so $\sigma P\subseteq P$.
	Since $\sigma$ is invertible, $\dim(\sigma P)=\dim P$;
	hence $\sigma P=P$.
	Finally, $\iota_i=\sigma^i\iota_0$ gives
	\[
	H_i=H\sigma^i\iota_0=T^iH_0,
	\qquad A_i=T^iA_0.
	\]
	
	\textup{(iii)}
	If $E=\{0\}$, then $P^\perp=\{0\}$, and both assertions hold
	with zero generators. Assume $E\ne\{0\}$, and let
	$E^*=\operatorname{Hom}_{\F_q}(E,\F_q)$ be its linear dual.
	For each $y\in P^\perp$, define a functional
	$\Phi(y)$ on $E$ by
	\[
	\Phi(y)(H(x))=\langle x,y\rangle
	\qquad\bigl(x\in(\F_q^a)^n\bigr).
	\]
	Since $H$ is onto, every element of $E$ has the form $H(x)$.
	To check that the definition is independent of the choice of $x$,
	suppose $H(x)=H(x')$. Then $x-x'\in\ker H=P$, and
	$y\in P^\perp$ gives
	\[
	\langle x,y\rangle-\langle x',y\rangle
	=\langle x-x',y\rangle=0.
	\]
	Thus $\Phi(y)$ is well defined. Linearity of $H$ and bilinearity
	of the pairing show that $\Phi(y)$ is a linear functional and
	that the map
	\[
	\Phi:P^\perp\longrightarrow E^*,\qquad y\longmapsto\Phi(y),
	\]
	is linear.
	
	If $\Phi(y)=0$, then $\langle x,y\rangle=0$ for every
	$x\in(\F_q^a)^n$. Nondegeneracy of the scalar pairing forces
	$y=0$, so $\Phi$ is injective.
	Finally, since $H$ is onto and $\ker H=P$,
	\[
	\dim P^\perp=\codim P=\dim E=\dim E^*.
	\]
	Hence $\Phi$ is an injective linear map between spaces of the
	same finite dimension, and is therefore an isomorphism.
	
	Since $\sigma P=P$, we also have $\sigma P^\perp=P^\perp$.
	Indeed, for $c\in P$ and $y\in P^\perp$,
	\[
	\langle c,\sigma y\rangle
	=\langle\sigma^{-1}c,y\rangle=0,
	\]
	because $\sigma^{-1}c\in P$. Thus $\sigma P^\perp\subseteq P^\perp$,
	and equality follows from invertibility of $\sigma$.
	
	Define the linear operator $S:E^*\to E^*$ by
	\[
	S(\lambda)=\lambda\circ T^{-1}.
	\]
	For $x\in(\F_q^a)^n$ and $y\in P^\perp$, we have
	\begin{equation}\label{eq:dual-action}
		\begin{aligned}
			\Phi(\sigma y)(H(x))
			=\langle x,\sigma y\rangle
			=\langle\sigma^{-1}x,y\rangle
			=\Phi(y)(T^{-1}H(x))
			=S(\Phi(y))(H(x)),
		\end{aligned}
	\end{equation}
	where the third equality uses $H\sigma^{-1}=T^{-1}H$.
	Since $H$ is onto, this proves
	\[
	\Phi\sigma=S\Phi.
	\]
	Thus shifting a word in $P^\perp$ corresponds, under $\Phi$,
	to applying $S$ to its associated functional.
	
	Choose a basis of $E$, and let $M$ be the matrix of $T$.
	Then $S$ has matrix $(M^{-1})^{\mathsf T}$ in the corresponding
	dual basis of $E^*$. A matrix and its transpose have the same
	minimal polynomial, so $S$ and $T^{-1}$ do as well.
	Recall that a linear operator has a cyclic vector exactly when its
	minimal polynomial has degree equal to the dimension of its space.
	Since $\dim E=\dim E^*$, it follows that $S$ has a cyclic vector
	if and only if $T^{-1}$ does.
	Moreover, $T^n=I_E$ gives
	\[
	\F_q[T]v=\F_q[T^{-1}]v
	\qquad(v\in E),
	\]
	so $T$ and $T^{-1}$ have the same cyclic vectors.
	Consequently, $T$ has a cyclic vector if and only if $S$ does.
	
	Finally, iterating $\Phi\sigma=S\Phi$ gives
	\[
	\Phi(\sigma^i z)=S^i\Phi(z)
	\qquad(z\in P^\perp,\ i\ge0).
	\]
	Because $\Phi$ is an isomorphism, we therefore have
	\[
	\Span_{\F_q}\{\sigma^i z:0\le i<n\}=P^\perp
	\quad\Longleftrightarrow\quad
	\Span_{\F_q}\{S^i\Phi(z):0\le i<n\}=E^*.
	\]
	Also $S^n=I_{E^*}$, since $T^n=I_E$.
	Thus the right-hand condition says precisely that $\Phi(z)$
	is a cyclic vector for $S$. Every functional in $E^*$ has the
	form $\Phi(z)$, so the existence of such a cyclic vector is
	equivalent to \eqref{eq:one-generator}.
	No square-freeness assumption on $X^n-1$ is needed.
\end{proof}

\begin{proof}[Proof of Theorem~\ref{thm:general-bridge}]
	\emph{From a covering to a code.}
	Suppose $h_q(n)\ge r$. Choose a covering of codimension at least
	$r$ and, if necessary, enlarge it to a covering $U\le V=\F_q^n$
	of codimension exactly $r$.
	Apply Lemma~\ref{lem:core-quotient} and use its notation
	$K,E,A,\pi,T$. Set
	\[
	d=\dim E=\codim K,\qquad a=\dim A=d-r.
	\]
	If $a=0$, then $U=K$ would be invariant and its shifts could not
	cover $V$, since $r>0$. Thus $a>0$, and $d\le n$ gives
	$a\le n-r$.
	
	Choose an $\F_q$-linear isomorphism $B:\F_q^a\to A$, and define
	\begin{equation}\label{eq:syndrome-construction}
		H:(\F_q^a)^n\longrightarrow E,\qquad
		H((x_i)_i)=\sum_{i=0}^{n-1}T^iB(x_i),\qquad P=\ker H.
	\end{equation}
	Since $U$ is cyclically covering,
	Lemma~\ref{lem:core-quotient} \textup{(iii)} gives
	$E=\bigcup_iT^iA$.
	Every $y\in E$ can therefore be written as $T^iB(b)$.
	It is the image under $H$ of the word with symbol $b$ in
	position $i$ and zeros elsewhere. Hence $H$ is onto, and
	\[
	\codim P=\dim E=d=r+a.
	\]
	
	Using $T^n=I_E$, we obtain
	\[
	\begin{aligned}
		H(\sigma x)
		=B(x_{n-1})+\sum_{i=1}^{n-1}T^iB(x_{i-1})
		=T\left(\sum_{j=0}^{n-1}T^jB(x_j)\right)=TH(x).
	\end{aligned}
	\]
	Here $H_i=T^iB$ is injective and $A_i=T^iA$.
	Lemma~\ref{lem:syndrome-criteria} \textup{(i)--(ii)} therefore gives
	cyclicity, the absence of nonzero words of block weight one,
	and $\rho(P)\le1$. Since $\codim P=r+a>0$, the code is proper,
	so $\rho(P)=1$.
	
	For this same operator $T$,
	Lemma~\ref{lem:core-quotient}\textup{(ii)} supplies the cyclic
	vector $\pi(e_0)$. Lemma~\ref{lem:syndrome-criteria}\textup{(iii)}
	then gives the dual condition \eqref{eq:one-generator}.
	
	\emph{From a code to a covering.}
	Conversely, let $P$ satisfy the stated conditions. Take the
	syndrome space $E=(\F_q^a)^n/P$ and its quotient map $H$, so
	$\dim E=\codim P=r+a$. Use $H_i,A_i$ and $T$ from
	Lemma~\ref{lem:syndrome-criteria}. Its first two parts give
	\[
	E=\bigcup_{i=0}^{n-1}A_i,\qquad
	\dim A_i=a,\qquad A_i=T^iA_0,\qquad T^n=I_E.
	\]
	By the dual condition and part \textup{(iii)}, choose a cyclic
	vector $v\in E$, and define the linear map
	\begin{equation}\label{eq:regular-pullback}
		\pi:\F_q^n\longrightarrow E,\qquad \pi(e_i)=T^iv,
	\end{equation}
	where $e_0,\ldots,e_{n-1}$ is the standard basis of $\F_q^n$.
	The cyclic-vector property makes $\pi$ onto. The identities
	$\tau e_i=e_{i+1}$, with indices modulo $n$, and $T^n=I_E$
	give $\pi\tau=T\pi$.
	
	Put $U=\pi^{-1}(A_0)$. Since $\pi\tau=T\pi$ and $T$ is
	invertible, a vector $w$ lies in $\tau^iU$ exactly when
	$\pi(w)\in T^iA_0$. Thus
	$\tau^iU=\pi^{-1}(T^iA_0)=\pi^{-1}(A_i)$.
	Pulling back the covering of $E$ now gives
	\[
	\bigcup_{i=0}^{n-1}\tau^iU
	=\pi^{-1}\left(\bigcup_{i=0}^{n-1}A_i\right)=\F_q^n,
	\qquad \codim U=\dim E-a=r.
	\]
	Intersecting the same identities gives
	$\core(U)=\pi^{-1}(\bigcap_iA_i)$.
	Finally, the code size is
	$|P|=q^{an-(r+a)}=q^{a(n-1)-r}$.
\end{proof}

We use the syndrome notation \eqref{eq:one-block-syndromes}
throughout the paper. Let
$H:(\F_q^a)^n\to E$ be an onto $\F_q$-linear map with
$\ker H=P$. The map
\[
(\F_q^a)^n/P\longrightarrow E,\qquad x+P\longmapsto H(x),
\]
is a linear isomorphism: each coset of $P$ corresponds to
exactly one syndrome. In chosen bases, $H$ is represented by
a $d\times an$ parity-check matrix of $P$, where $d=\dim E$.
This matrix has full row rank because $H$ is onto.

For a cyclically covering subspace $U\le\F_q^n$ of positive
codimension, we call a code obtained by
\eqref{eq:syndrome-construction} a \emph{normalized syndrome code}
of $U$. Its syndrome space is the core quotient
$\F_q^n/\core(U)$, and its parameters satisfy
\begin{equation}\label{eq:core-parameters}
	r=\codim U,\qquad d=\codim\core(U),\qquad a=d-r.
\end{equation}
Moreover, \eqref{eq:core-quotient-cover} gives
$\bigcap_{i=0}^{n-1}A_i=\{0\}$ for its one-block syndrome spaces.

The construction involves choosing an isomorphism
$B:\F_q^a\to U/\core(U)$.
Different choices of $B$ give codes related by the same
invertible linear change of alphabet coordinates in every block.
Their scalar duals are related by the inverse transpose of
that change, again applied in every block.
Both transformations preserve block weight and commute with
the block shift $\sigma$.
Consequently, cyclicity and the one-generator condition
\eqref{eq:one-generator} do not depend on the choice of $B$
or the coordinate bases.

\begin{example}\label{ex:binary-syndromes}
	Let $V=\F_2^3$ and
	\[
	U=\{x\in V:x_0=x_1\}.
	\]
	Every binary triple has two equal coordinates, so the three
	cyclic shifts of $U$ cover $V$. A vector belongs to all three
	shifts precisely when all its coordinates are equal. Hence
	\[
	K=\core(U)=\langle(1,1,1)\rangle.
	\]
	Thus
	\[
	r=\codim U=1,\qquad d=\codim K=2,\qquad a=d-r=1,
	\]
	so the associated code has alphabet $\F_2$.
	
	We now construct this code explicitly.
	Let $e_0,e_1,e_2$ be the standard basis of $V$, and choose
	\[
	(e_2+K,e_0+K)
	\]
	as an ordered basis of the syndrome space $E=V/K$.
	Since
	\[
	\tau e_2=e_0,
	\qquad
	\tau e_0=e_1\equiv e_0+e_2\pmod K,
	\]
	the induced shift on $E$ has matrix
	\[
	T=\begin{pmatrix}0&1\\1&1\end{pmatrix}.
	\]
	Moreover,
	\[
	U=\Span_{\F_2}\{e_0+e_1,e_2\},
	\qquad
	e_0+e_1+K=e_2+K.
	\]
	Therefore $A=U/K$ is spanned by $e_2+K$. In the chosen
	coordinates on $E$, this means
	\[
	A=\Span_{\F_2}\{b\},
	\qquad b=\binom10.
	\]
	
	Let $B:\F_2\to A$ be the linear isomorphism defined by
	$B(t)=tb$. The syndrome construction gives
	\[
	H(x_0,x_1,x_2)
	=\sum_{i=0}^{2}T^iB(x_i)
	=x_0b+x_1Tb+x_2T^2b.
	\]
	Thus the matrix of $H$ is
	\[
	[\,b\;\;Tb\;\;T^2b\,]
	=\begin{pmatrix}1&0&1\\0&1&1\end{pmatrix}.
	\]
	In particular, $H(x_0,x_1,x_2)=0$ if and only if
	$x_0=x_1=x_2$. Hence
	\[
	P=\ker H=\{000,111\},
	\]
	the binary repetition code of length three, with redundancy
	$r+a=2$.
	
	Finally, the four errors of weight at most one,
	\[
	000,\qquad 100,\qquad 010,\qquad 001,
	\]
	have respective syndromes
	\[
	0,\qquad b,\qquad Tb,\qquad T^2b.
	\]
	These are precisely the four elements of $E$.
	Consequently, for every received word $x\in\F_2^3$, there is
	a unique error $e$ of weight at most one such that $H(e)=H(x)$.
	Then $H(x-e)=0$, so $x-e\in P$.
	Thus every word in $\F_2^3$ differs from a unique codeword
	of $P$ in at most one coordinate.
\end{example}

The alphabet dimension of the normalized syndrome code is
$a=\dim U-\dim\core(U)$. Even optimal cyclic coverings of the
same dimension can therefore give codes over vector alphabets
of different dimensions.

\begin{example}\label{ex:alphabet-dimension-dependence}
	Let $V=\F_2^7$, with indices modulo seven, and put
	\[
	\begin{aligned}
		U_1&=\{x\in V:x_0=x_1=x_3\},\\
		U_2&=\{x\in V:x_4+x_6=x_1+x_5=x_2+x_3\}.
	\end{aligned}
	\]
	Both are optimal cyclically covering subspaces of dimension five,
	with $h_2(7)=2$. However,
	\[
	\dim\core(U_1)=1,\qquad \dim\core(U_2)=4.
	\]
	Their normalized syndrome codes therefore have alphabet
	dimensions $a_1=4$ and $a_2=1$, respectively.
\end{example}

\begin{proof}
	Each subspace is defined by two independent linear constraints,
	so $\dim U_1=\dim U_2=5$.
	
	To prove that the shifts of $U_1$ cover $V$, consider the seven
	triples
	\[
	L_i=\{i,i+1,i+3\},\qquad i\in\mathbb Z/7\mathbb Z.
	\]
	Every unordered pair of indices belongs to exactly one $L_i$:
	the six differences between distinct elements of $\{0,1,3\}$
	are precisely the nonzero residues modulo seven.
	
	Suppose that a binary word $x$ lies in none of the shifts of $U_1$.
	Then no $L_i$ has three equal coordinates, so each $L_i$ contains
	exactly one pair of equal coordinates.
	There would therefore be exactly seven equal-coordinate pairs.
	On the other hand, if $x$ has $w$ coordinates equal to one, their
	total number is
	\[
	\binom{w}{2}+\binom{7-w}{2}\ge9,
	\]
	a contradiction. Thus $U_1$ is cyclically covering. Its cyclic
	core is
	\[
	\core(U_1)=\langle\mathbf1\rangle,
	\qquad \mathbf1=(1,1,1,1,1,1,1),
	\]
	because the shifted defining constraints include
	$x_i=x_{i+1}$ for every $i$. Hence $\dim\core(U_1)=1$.
	
	Now put $C=I+\tau+\tau^3$. Since
	$(Cx)_i=x_i+x_{i-1}+x_{i-3}$, the condition
	$(Cx)_0=(Cx)_1=(Cx)_3$ is precisely the defining condition for
	$U_2$. Thus $U_2=C^{-1}(U_1)$, where $C^{-1}(U_1)$ denotes
	the inverse image. Because $C\tau=\tau C$,
	\[
	\bigcup_{i=0}^{6}\tau^iU_2
	=C^{-1}\left(\bigcup_{i=0}^{6}\tau^iU_1\right)=V.
	\]
	Hence $U_2$ is also cyclically covering, and similarly
	\[
	\core(U_2)=C^{-1}(\core(U_1))
	=C^{-1}(\langle\mathbf1\rangle).
	\]
	
	Solving $Cx=0$ gives
	\[
	\ker C=
	\{(s,t,u,s+u,s+t+u,s+t,t+u):s,t,u\in\F_2\},
	\]
	so $\dim\ker C=3$. Since $C\mathbf1=\mathbf1$, we obtain
	\[
	\core(U_2)=\ker C\oplus\langle\mathbf1\rangle,
	\]
	and therefore $\dim\core(U_2)=4$.
	
	Finally, the counting bound gives
	\[
	h_2(7)\le \lfloor\log_2 7\rfloor=2.
	\]
	Both coverings have codimension two, so $h_2(7)=2$ and both are
	optimal. Their alphabet dimensions are $a_1=5-1=4$ and
	$a_2=5-4=1$.
\end{proof}

\begin{definition}
	\label{def:uniform-alphabet-equivalence}
	Let $P\le(\F_q^a)^n$ and $P'\le(\F_q^{a'})^n$.
	We say that $P$ and $P'$ are \emph{equivalent under a uniform
		change of alphabet coordinates} if $a=a'$ and
	$P'=D_S(P)$ for some $S\in\GL(a,q)$, where
	\[
	D_S(x_0,\ldots,x_{n-1})=(Sx_0,\ldots,Sx_{n-1}).
	\]
	Thus the same invertible linear map is applied in every block.
\end{definition}

The map $D_S$ preserves block Hamming distance and commutes with
the block shift $\sigma$. In the following proposition,
equivalence always means equivalence under a uniform change of
alphabet coordinates.

\begin{proposition}\label{prop:uniform-syndrome-equivalence}
	Let $U,U'\le V=\F_q^n$ be cyclically covering subspaces of positive
	codimension, and let $P,P'$ be their normalized syndrome codes.
	Then $P$ and $P'$ are equivalent if and only if there is an
	invertible linear map $C:V\to V$ such that
	\[
	C\tau=\tau C,\qquad U'=C(U).
	\]
	Equivalently, they are equivalent if and only if
	\[
	U'=c(\tau)U
	\quad\text{for some }c\in\F_q[X]
	\text{ with }\gcd(c,X^n-1)=1.
	\]
	Whenever these conditions hold, $\core(U)=\core(U')$.
\end{proposition}

\begin{proof}
	Put $K=\core(U)$, $E=V/K$, and $A=U/K$, and let $T$ be the
	operator induced by $\tau$ on $E$.
	Choose an isomorphism $B:\F_q^a\to A$. The normalized syndrome
	map is
	\[
	H(x)=\sum_{i=0}^{n-1}T^iB(x_i),\qquad P=\ker H.
	\]
	Use primed notation for $U'$ and $P'$.
	
	Identify $V$ with $R=\F_q[X]/(f)$, where $f=X^n-1$, by
	$e_i\leftrightarrow X^i$. Then $\tau$ acts as multiplication by
	$X$. Since $e_0$ is a cyclic vector for $\tau$, every linear map
	commuting with $\tau$ has the form $c(\tau)$; it is invertible
	exactly when $\gcd(c,f)=1$.
	
	Suppose first that $U'=C(U)$, where $C=c(\tau)$ is invertible.
	Since $C$ commutes with $\tau$,
	\[
	K'=\bigcap_i\tau^iCU
	=C\bigcap_i\tau^iU
	=CK.
	\]
	The cyclic subspace $K$ is an ideal of $R$, so multiplication by
	the unit $c$ preserves $K$. Thus $K'=K$.
	The map $C$ therefore induces an isomorphism $F:E\to E'$ with
	\[
	FT=T'F,\qquad F(A)=A'.
	\]
	In particular, $a=a'$, and
	$S=(B')^{-1}FB$ belongs to $\GL(a,q)$. For every $x$,
	\[
	\begin{aligned}
		H'D_S(x)
		&=\sum_i(T')^iB'S(x_i)\\
		&=\sum_i(T')^iFB(x_i)\\
		&=F\sum_iT^iB(x_i)
		=FH(x).
	\end{aligned}
	\]
	Taking kernels gives $P'=D_S(P)$.
	
	Conversely, suppose that $P'=D_S(P)$.
	The maps $H$ and $H'D_S$ are onto and have the same kernel,
	so there is an isomorphism $F:E\to E'$ satisfying
	\[
	FH=H'D_S.
	\]
	The identities
	\[
	H\sigma=TH,\qquad
	H'\sigma=T'H',\qquad
	D_S\sigma=\sigma D_S
	\]
	imply $FT=T'F$. Restricting $FH=H'D_S$ to the zeroth block
	also gives $FB=B'S$, and hence $F(A)=A'$.
	
	Thus $F$ is an $R$-module isomorphism $R/K\to R/K'$.
	The elements of $R$ annihilating every element of $R/K$ form
	exactly the ideal $K$, and similarly for $K'$.
	An $R$-module isomorphism preserves these annihilators, so
	$K=K'$.
	
	Write $K=(g)$, where $g$ is the unique monic divisor of $f$
	generating this ideal. Under the identification
	$E\cong\F_q[X]/(g)$, the map $F$ is multiplication by a unit
	$u$ modulo $g$, since an $R$-linear map is determined by its
	value at $1$.
	
	To lift this unit modulo $f$, let $h$ be the product of the full
	prime-power factors of $f$ whose irreducible factors do not divide
	$g$, with $h=1$ if there are none.
	Since $\gcd(g,h)=1$, the Chinese remainder theorem gives a
	polynomial $c$ such that
	\[
	c\equiv u\pmod g,\qquad c\equiv1\pmod h.
	\]
	Every irreducible divisor of $f$ divides either $g$ or $h$, so
	$\gcd(c,f)=1$. Hence $C=c(\tau)$ is invertible and induces $F$
	on $V/K$. Since $U$ and $U'$ are the inverse images of $A$ and
	$A'$, respectively, the identity $F(A)=A'$ yields $C(U)=U'$.
\end{proof}

\section{Cyclic perfect codes and arithmetic criteria}\label{sec:perfect}
A code $P\le(\F_q^a)^n$ is \emph{perfect} if every
$x\in(\F_q^a)^n$ has a unique expression $x=c+e$ with
$c\in P$ and $\wt(e)\le1$.
Equivalently, its radius-one balls partition the ambient space.

The one-block syndrome spaces from Section~\ref{sec:structure}
give a geometric criterion for this uniqueness.

\begin{lemma}\label{lem:perfect-syndromes}
Let $H:(\F_q^a)^n\to E$ be an onto linear map with kernel $P$.
Write $H_i=H\iota_i$ and $A_i=\im H_i$.
The code $P$ is perfect if and only if
\[
 E=\bigcup_{i=0}^{n-1}A_i,\qquad
 \dim A_i=a\quad\text{for every }i,\qquad
 A_i\cap A_j=\{0\}\quad\text{for }i\ne j.
\]
\end{lemma}

\begin{proof}
	Since $H_i$ has domain $\F_q^a$ and image $A_i$,
	it is injective exactly when $\dim A_i=a$.
	
	To prove the perfectness criterion, observe that for
	$x,e\in(\F_q^a)^n$,
	\[
	H(x)=H(e)
	\quad\Longleftrightarrow\quad
	x-e\in\ker H=P.
	\]
	Thus choosing an error $e$ with the same syndrome as $x$
	determines the codeword $c=x-e$ uniquely.
	Since $H$ is onto, $P$ is perfect precisely when every
	$y\in E$ is the syndrome of exactly one error of block weight
	at most one.
	
	Suppose the three stated conditions hold.
	The equality $E=\bigcup_iA_i$ ensures that every syndrome
	has an error representative of weight at most one.
	The condition $\dim A_i=a$ makes each $H_i$ injective,
	so the zero syndrome has only the zero error as such a
	representative.
	For a nonzero syndrome $y$, suppose
	\[
	y=H_i(b)=H_j(b'),
	\qquad b,b'\ne0.
	\]
	If $i\ne j$, then $y\in A_i\cap A_j$, contradicting the
	trivial-intersection condition.
	If $i=j$, injectivity gives $b=b'$.
	Hence the error representative is unique, and $P$ is perfect.
	
	Conversely, suppose $P$ is perfect.
	Every syndrome has an error representative of weight at most
	one, so $E=\bigcup_iA_i$.
	If some $H_i$ were not injective, a nonzero error in position
	$i$ would have the same syndrome as the zero error.
	Thus every $H_i$ is injective, giving $\dim A_i=a$.
	Finally, a nonzero vector in $A_i\cap A_j$ with $i\ne j$
	would be the syndrome of two distinct nonzero errors supported
	in different positions.
	Uniqueness excludes this possibility, so
	$A_i\cap A_j=\{0\}$.
\end{proof}

The perfect-syndrome criterion can also be expressed directly
in terms of intersections of the shifted covering subspaces.

\begin{corollary}\label{cor:perfect-cyclic-intersections}
	Let $U\le\F_q^n$ be a cyclically covering subspace of positive
	codimension, and let $P$ be its normalized syndrome code. Then
	\begin{equation}\label{eq:perfect-cyclic-intersections}
		P\text{ is perfect}
		\quad\Longleftrightarrow\quad
		U\cap\tau^jU=\core(U)
		\quad\text{for every }1\le j<n.
	\end{equation}
\end{corollary}

\begin{proof}
	Put $K=\core(U)$, $E=\F_q^n/K$, and $A=U/K$, and let $T$
	be the operator induced by $\tau$ on $E$.
	The one-block syndrome maps of the normalized construction are
	$H_i=T^iB$. Each $H_i$ is injective, and their images
	$A_i=T^iA$ cover $E$.
	By Lemma~\ref{lem:perfect-syndromes}, the code $P$ is therefore
	perfect if and only if
	\[
	A_i\cap A_j=\{0\}\qquad(i\ne j).
	\]
	Since $T$ is invertible, this is equivalent to
	\[
	A\cap T^jA=\{0\}\qquad(1\le j<n).
	\]
	Moreover, $K\subseteq U\cap\tau^jU$, so
	\[
	A\cap T^jA
	=(U/K)\cap(\tau^jU/K)
	=(U\cap\tau^jU)/K.
	\]
	Consequently, $A\cap T^jA=\{0\}$ if and only if
	$U\cap\tau^jU=K$, proving the assertion.
\end{proof}

Counting the nonzero syndromes gives a general bound for coverings
and the parameter restrictions imposed by perfectness.

\begin{lemma}\label{lem:sphere-capacity}
	Let $U\le\F_q^n$ be a cyclically covering subspace of codimension
	$r>0$, and let $P\le(\F_q^a)^n$ be its normalized syndrome code.
	Then
	\begin{equation}\label{eq:sphere-capacity}
		n(q^a-1)\ge q^{r+a}-1,
	\end{equation}
	with equality if and only if $P$ is perfect.
	
	If $P$ is perfect, then
	\begin{equation}\label{eq:lengths}
		n=\frac{q^{r+a}-1}{q^a-1},\qquad a\mid r.
	\end{equation}
	Consequently, its parameters can be written as
	\begin{equation}\label{eq:general-hamming}
		a=k,\qquad D=km,\qquad r=k(m-1),\qquad
		n=N_{q,k,m}=\frac{q^{km}-1}{q^k-1},
		\qquad k\ge1,\quad m\ge2,
	\end{equation}
	where $D$ is the syndrome dimension.
	At these parameters, $q^r<n<q^{r+1}$, so every
	codimension-$r$ cyclic covering attains the counting bound.
\end{lemma}

\begin{proof}
	Let $E=\F_q^n/\core(U)$ be the syndrome space of $P$.
	For each $0\le i<n$, write $A_i=\im H_i$, where
	$H_i:\F_q^a\to E$ is the syndrome map for errors supported
	in position $i$.
	The normalized construction gives
	\[
	\dim E=r+a,\qquad
	\dim A_i=a,\qquad
	E=\bigcup_{i=0}^{n-1}A_i,
	\]
	and every $H_i$ is injective.
	
	Each $A_i$ contains $q^a-1$ nonzero syndromes.
	Since these sets cover all nonzero elements of $E$,
	\[
	q^{r+a}-1
	=\left|\bigcup_{i=0}^{n-1}(A_i\setminus\{0\})\right|
	\le\sum_{i=0}^{n-1}|A_i\setminus\{0\}|
	=n(q^a-1).
	\]
	This proves \eqref{eq:sphere-capacity}.
	Equality holds exactly when every nonzero syndrome is counted
	once, or equivalently when
	$A_i\cap A_j=\{0\}$ for all $i\ne j$.
	Together with the injectivity of the one-block maps,
	Lemma~\ref{lem:perfect-syndromes} shows that this is precisely
	the condition that $P$ is perfect.
	
	Suppose now that $P$ is perfect.
	Solving the equality in \eqref{eq:sphere-capacity} for $n$
	gives the length formula in \eqref{eq:lengths}.
	Its integrality implies $q^a-1\mid q^{r+a}-1$, so
	\[
	q^a-1
	=\gcd(q^a-1,q^{r+a}-1)
	=q^{\gcd(a,r+a)}-1.
	\]
	Comparing exponents gives $\gcd(a,r+a)=a$, hence $a\mid r$.
	We may therefore set
	\[
	k=a,\qquad m=1+\frac ra.
	\]
	Since $a\ge1$ and $r>0$, these are integers with
	$k\ge1$ and $m\ge2$.
	Moreover,
	\[
	r=k(m-1),\qquad D=\dim E=r+a=km,
	\]
	which yields the parameters in \eqref{eq:general-hamming}.
	
	Finally, these parameters alone determine the counting bound.
	Indeed,
	\[
	n=1+q^k+\cdots+q^{k(m-1)}.
	\]
	The last term is $q^r$, and the other terms are positive,
	so $n>q^r$.
	The exponents $0,k,\ldots,k(m-1)$ are distinct integers
	between $0$ and $r$, giving
	\[
	n\le\sum_{j=0}^{r}q^j
	=\frac{q^{r+1}-1}{q-1}
	<q^{r+1}.
	\]
	Thus
	\[
	h_q(n)\le\lfloor\log_q n\rfloor=r.
	\]
	Any codimension-$r$ cyclic covering at this length gives
	$h_q(n)\ge r$.
	Hence every such covering attains the counting bound.
\end{proof}

We now ask when the parameters in \eqref{eq:general-hamming}
admit a cyclic perfect code.
The length formula alone does not guarantee cyclicity.
Let $I_q(j)$ denote the number of monic irreducible polynomials
of degree $j$ over $\F_q$, and put
\[
 S_{q,k}=\sum_{j=1}^{k-1}I_q(j).
\]
The following arithmetic criterion applies in prime alphabet
dimension and, for arbitrary alphabet dimension, above an
explicit extension-degree bound.

\begin{theorem}\label{thm:prime-block}
Let $q$ be a prime power, let $k,m\ge2$, and put
\[
 n=N_{q,k,m}=\frac{q^{km}-1}{q^k-1},\qquad
 r=k(m-1),\qquad D=km.
\]
Suppose that $k$ is prime or $m\ge2S_{q,k}+2$.
Then a cyclic $\F_q$-linear perfect code of length $n$ over
$\F_q^k$ exists if and only if
\begin{equation}\label{eq:prime-criterion}
 \gcd(m,q-1)=1.
\end{equation}
When this condition holds, there is a codimension-$r$ cyclic
covering whose cyclic core has codimension $D$, and
$h_q(n)=r$.
\end{theorem}

For $k=2$, related cyclic subspace constructions appear in
\cite{PauleyBamberg,FengLu,JamesonSheekey}, under their respective
parameter hypotheses. The theorem gives a uniform criterion for
every prime $k$ and, for each fixed $q,k$, for all sufficiently
large $m$.

\begin{proof}[Proof of necessity in Theorem~\ref{thm:prime-block}]
	Let $P\le(\F_q^k)^n$ be cyclic and perfect, and let
	\[
	W=(\F_q^k)^n/P
	\]
	be its syndrome space.
	For each $0\le i<n$, let $A_i\le W$ be the image, under the
	quotient map, of the words supported in position $i$.
	By Lemma~\ref{lem:perfect-syndromes}, each $A_i$ has dimension
	$k$, and every nonzero vector of $W$ belongs to exactly one
	$A_i$. In particular, these subspaces are distinct.
	
	The cyclic shift induces an operator $T$ on $W$.
	Lemma~\ref{lem:syndrome-criteria}\textup{(ii)} gives
	\[
	T^n=I_W,\qquad A_i=T^iA_0.
	\]
	We show that a common prime divisor of $m$ and $q-1$ would
	force two distinct subspaces $A_i$ to share a nonzero vector.
	
	Suppose, for a contradiction, that $\gcd(m,q-1)>1$.
	Since
	\[
	n=1+q^k+\cdots+q^{k(m-1)}\equiv m\pmod{q-1},
	\]
	there is a prime $\ell$ dividing both $n$ and $q-1$.
	Put $t=n/\ell$ and $S=T^t$. Then
	\[
	0<t<n,\qquad S^\ell=I_W.
	\]
	Because $\ell\mid q-1$, the polynomial $X^\ell-1$ splits
	into distinct linear factors over $\F_q$.
	It annihilates $S$, so $S$ is diagonalizable over $\F_q$.
	We may therefore choose a nonzero eigenvector $x\in W$ with
	\[
	Sx=\lambda x,\qquad \lambda\in\F_q^\times.
	\]
	
	Let $A_i$ be the unique subspace containing $x$.
	Since $A_i$ is an $\F_q$-subspace, it also contains $\lambda x$.
	On the other hand,
	\[
	Sx\in SA_i=T^{i+t}A_0=A_{i+t},
	\]
	where indices are taken modulo $n$. Hence
	\[
	0\ne Sx=\lambda x\in A_i\cap A_{i+t}.
	\]
	Since $0<t<n$, these are distinct subspaces, contradicting
	the uniqueness of the subspace containing a nonzero syndrome.
	Thus $\gcd(m,q-1)=1$.
	The same argument applies to every $k\ge1$.
\end{proof}

For sufficiency, put $E=\F_{q^{km}}$ and $K=\F_{q^k}$.
The field norm is
\[
 \Nrm_{E/K}(x)=\prod_{j=0}^{m-1}x^{q^{kj}}=x^n.
\]
We construct a $k$-dimensional $\F_q$-subspace $A\le E$ on which
this norm restricts to a bijection $A\to K$.
The next two lemmas supply the spaces needed for the two
parameter ranges.

\begin{lemma}\label{lem:binomial-general}
Let $k\ge2$, put
\[
 L_{q,k}=\operatorname{lcm}_{1\le j<k}(q^j-1),
\]
and let $s>1$ be odd.
Suppose every prime divisor of $s$ divides $q^k-1$ and
$\gcd(s,L_{q,k})=1$.
There are a primitive element $c\in K=\F_{q^k}$ and an element
$\theta\in\F_{q^{ks}}$ with minimal polynomial $X^s-c$ over $K$
such that
\[
 A=\{f(\theta):f\in\F_q[X],\ \deg f<k\}
\]
has dimension $k$ over $\F_q$, and the norm from $\F_{q^{ks}}$
to $K$ restricts to a bijection $A\to K$.
The zero polynomial is included.
\end{lemma}

\begin{proof}
	Put $Q=q^k$ and $L=L_{q,k}$, and let $\mathcal V$ be the
	$k$-dimensional $\F_q$-space of polynomials of degree less than
	$k$, including the zero polynomial.
	
	We first choose the evaluation element $\theta$.
	For every prime $\ell\mid s$, the assumptions give
	$\ell$ odd and $\ell\mid Q-1$. Thus, for every $j\ge1$,
	\[
	v_\ell(Q^j-1)=v_\ell(Q-1)+v_\ell(j),
	\]
	where $v_\ell(t)$ denotes the exponent of $\ell$ in $t$.
	Since $Q-1$ already divides $Q^j-1$, divisibility by
	$s(Q-1)$ is equivalent to
	$v_\ell(j)\ge v_\ell(s)$ for every prime $\ell\mid s$.
	Hence
	\[
	s(Q-1)\mid Q^j-1
	\quad\Longleftrightarrow\quad s\mid j.
	\]
	Taking $j=s$, choose $\theta\in\F_{Q^s}^{\times}$ of order
	$s(Q-1)$ and put $c=\theta^s$.
	Then $c$ has order $Q-1$, so it is a primitive element of $K$.
	
	The degree of $\theta$ over $K$ is the least positive $j$
	such that $\theta^{Q^j}=\theta$.
	By the divisibility criterion above, this degree is $s$.
	Since $\theta^s=c$, its minimal polynomial over $K$ is
	therefore $X^s-c$.
	Moreover,
	\[
	K=\F_q(c)\subseteq\F_q(\theta),
	\]
	so $\F_q(\theta)=K(\theta)$ has degree $ks$ over $\F_q$.
	Consequently, evaluation at $\theta$ is injective on
	$\mathcal V$. Its image is $A$, giving $\dim_{\F_q}A=k$.
	
	We next construct a permutation of $\mathcal V$.
	For a nonzero polynomial
	\[
	f(X)=b\prod_{j=1}^t(X-\alpha_j)\in\mathcal V,
	\qquad b\in\F_q^\times,
	\]
	define
	\[
	\mathcal R_sf(X)=b^s\prod_{j=1}^t(X-\alpha_j^s),
	\qquad \mathcal R_s0=0.
	\]
	Roots are counted with multiplicity; for a constant polynomial,
	the product is empty.
	The Frobenius map $x\mapsto x^q$ permutes the roots of $f$
	and commutes with taking $s$th powers.
	Thus the transformed root multiset is again Frobenius-stable,
	so $\mathcal R_sf$ has coefficients in $\F_q$.
	Its degree equals that of $f$, and hence $\mathcal R_s$ maps
	$\mathcal V$ into itself.
	
	Because $\gcd(s,L)=1$, choose a positive integer $e$ with
	\[
	se\equiv1\pmod L.
	\]
	Each nonzero root $\alpha_j$ has degree $h<k$ over $\F_q$,
	so its multiplicative order divides $q^h-1$, and hence $L$.
	Also $q-1\mid L$. Therefore
	\[
	\alpha_j^{se}=\alpha_j,\qquad b^{se}=b.
	\]
	Positive powers also fix a zero root.
	Define $\mathcal R_e$ by the same formula with $s$ replaced by
	$e$. Applying the two transformations successively restores
	every root, with its multiplicity, and the leading coefficient:
	\[
	\mathcal R_e(\mathcal R_sf)
	=\mathcal R_s(\mathcal R_ef)=f
	\qquad(f\in\mathcal V).
	\]
	This includes constants and the zero polynomial.
	Thus $\mathcal R_s$ is a permutation of $\mathcal V$.
	
	Finally, we compute the norm of $f(\theta)$.
	The $K$-conjugates of $\theta$ are precisely the roots of
	$X^s-c$. Since $s$ is odd, for every root $\alpha$ of $f$,
	\[
	\prod_{\zeta^s=c}(\zeta-\alpha)
	=(-1)^s(\alpha^s-c)
	=c-\alpha^s.
	\]
	It follows that
	\begin{equation}\label{eq:norm-resultant}
		\begin{aligned}
			\Nrm_{\F_{Q^s}/K}(f(\theta))
			=\prod_{\zeta^s=c}f(\zeta)
			=b^s\prod_{j=1}^t(c-\alpha_j^s)
			=(\mathcal R_sf)(c).
		\end{aligned}
	\end{equation}
	The identity also holds for $f=0$.
	
	Since $c$ has degree $k$ over $\F_q$, evaluation at $c$
	bijects $\mathcal V$ with $K$.
	Together with the bijection $f\mapsto f(\theta)$ from
	$\mathcal V$ to $A$, formula \eqref{eq:norm-resultant}
	identifies the restricted norm with the composition
	\[
	A\xrightarrow{\,f(\theta)\mapsto f\,}\mathcal V
	\xrightarrow{\,\mathcal R_s\,}\mathcal V
	\xrightarrow{\,g\mapsto g(c)\,}K.
	\]
	All three maps are bijective, proving the claim.
	The argument applies to every $\theta$ of order $s(Q-1)$.
\end{proof}

\begin{lemma}\label{lem:large-extension}
Let $k\ge2$ and $m\ge2S_{q,k}+2$, and suppose that
$\gcd(m,q-1)=1$.
Put $E=\F_{q^{km}}$ and $K=\F_{q^k}$.
There is a $k$-dimensional $\F_q$-subspace $A\le E$ on which
$\Nrm_{E/K}$ restricts to a bijection $A\to K$.
\end{lemma}

\begin{proof}
	Put $Q=q^k$ and let $\mathcal P$ be the set of monic
	irreducible polynomials over $\F_q$ of degrees $1,\ldots,k-1$.
	Write
	\[
	S=|\mathcal P|=S_{q,k},\qquad
	B=\sum_{P\in\mathcal P}\deg P,\qquad
	h=(Q-1)^S,\qquad Z=Q^m.
	\]
	Counting roots in $\F_{q^j}$ gives $jI_q(j)\le q^j$, so
	\[
	B\le\sum_{j=1}^{k-1}q^j<Q-1,\qquad h<Q^S.
	\]
	Choose $c\in K$ of degree $k$ over $\F_q$.
	Then $P(c)\ne0$ for every $P\in\mathcal P$.
	We show that there is an element $\theta\in E$ satisfying
	\[
	\Nrm_{E/K}(P(\theta))=P(c)
	\qquad(P\in\mathcal P).
	\]

	Let $\Omega$ consist of the elements of $E$ at which every
	$P\in\mathcal P$ is nonzero, so $|\Omega|\ge Z-B$.
	Let $C_m$ count the elements of $\Omega$ satisfying the
	displayed norm conditions.
	If $\widehat{K^\times}$ denotes the group of multiplicative
	characters of $K^\times$, character orthogonality gives
	\[
	C_m=\frac1h\sum_{\theta\in\Omega}
	\prod_{P\in\mathcal P}
	\left(\sum_{\chi_P\in\widehat{K^\times}}
	\chi_P\!\left(\frac{\Nrm_{E/K}(P(\theta))}{P(c)}\right)\right).
	\]
	After expanding the product, the choice of all trivial
	characters contributes $|\Omega|/h$.
	For any other choice, let $b$ be the sum of the degrees of
	the polynomials whose characters are nontrivial.
	Each such character remains nontrivial after composition
	with the surjective norm $E^\times\to K^\times$.
	The polynomials in this support remain pairwise coprime and
	squarefree over $E$.
	The mixed character-sum bound of
	Wan~\cite[Corollary~2.3]{WanGenerators} therefore bounds the
	complete sum over $E$ by $(b-1)\sqrt Z$ in absolute value;
	here nontrivial characters are extended by zero at zero.
	Deleting the roots of the remaining polynomials changes
	the sum by at most $B-b$.
	The constant factors involving $P(c)$ have absolute value one.
	Consequently, each of the $h-1$ nonprincipal sums has
	absolute value at most
	\[
	(b-1)\sqrt Z+B-b\le(B-1)\sqrt Z,
	\]
	and hence
	\[
	C_m\ge\frac{Z-B-(h-1)(B-1)\sqrt Z}{h}.
	\]

	Put $t=\sqrt Z$ and $c_0=(h-1)(B-1)$.
	Our assumption on $m$ gives $t\ge Q^{S+1}$, whereas the
	integer $c_0$ satisfies $c_0<Q^{S+1}$.
	Thus $t-c_0\ge1$ and
	\[
	Z-c_0\sqrt Z-B=t(t-c_0)-B\ge t-B>0.
	\]
	It follows that $C_m>0$.

	Choose a counted element $\theta$, and set
	\[
	A=\{f(\theta):f\in\F_q[X],\ \deg f<k\},
	\]
	including the zero polynomial.
	No nonzero polynomial of degree less than $k$ vanishes at
	$\theta$, since one of its irreducible factors would belong
	to $\mathcal P$, contradicting $\theta\in\Omega$.
	Thus $\dim_{\F_q}A=k$.
	For a nonzero $f=ag$, where $a\in\F_q^\times$ and $g$ is
	monic of degree less than $k$, factor $g$ into members of
	$\mathcal P$. The norm conditions give
	\[
	\Nrm_{E/K}(f(\theta))=a^m g(c).
	\]
	Since $\gcd(m,q-1)=1$, the map $ag\mapsto a^m g$, fixing
	zero, permutes the polynomials of degree less than $k$.
	Evaluation at $c$ bijects these polynomials with $K$.
	Therefore the norm restricts to a bijection $A\to K$.
\end{proof}

\begin{proof}[Proof of sufficiency in Theorem~\ref{thm:prime-block}]
	Assume $\gcd(m,q-1)=1$, and put
	$Q=q^k$, $K=\F_Q$, and $E=\F_{Q^m}$.
	We first construct a $k$-dimensional $\F_q$-subspace of $E$
	on which the norm to $K$ is bijective.
	
	Suppose first that $k$ is prime.
	Write $m=su$, where $s$ is the product of the full prime-power
	factors of $m$ whose primes divide $Q-1$.
	Then $\gcd(u,Q-1)=1$.
	For every prime $\ell\mid s$, we have $\ell\mid q^k-1$.
	Since $\ell\mid m$ and $\gcd(m,q-1)=1$, we also have
	$\ell\nmid q-1$.
	Thus $\ord_\ell(q)$ divides the prime $k$ but is not one,
	so $\ord_\ell(q)=k$.
	In particular, $\ell$ is odd and divides none of $q^j-1$
	for $1\le j<k$.
	Hence $s$ is odd and $\gcd(s,L_{q,k})=1$.
	
	Put $E_0=\F_{Q^s}\subseteq E$.
	If $s=1$, take $A=K=E_0$, whose norm to $K$ is the identity.
	If $s>1$, Lemma~\ref{lem:binomial-general} supplies a
	$k$-dimensional $\F_q$-subspace $A\le E_0$ on which
	$\Nrm_{E_0/K}$ is bijective onto $K$.
	Since $[E:E_0]=u$, every $x\in E_0$ satisfies
	$\Nrm_{E/E_0}(x)=x^u$.
	Norm transitivity therefore gives
	\[
	\Nrm_{E/K}(x)
	=\Nrm_{E_0/K}(x^u)
	=\Nrm_{E_0/K}(x)^u.
	\]
	The power map $y\mapsto y^u$ permutes $K^\times$ and fixes
	zero, because $\gcd(u,Q-1)=1$.
	Thus, in both cases,
	$\Nrm_{E/K}|_A:A\to K$ is bijective.
	
	If $k$ is not prime, the hypothesis of the theorem gives
	$m\ge2S_{q,k}+2$, and Lemma~\ref{lem:large-extension}
	supplies such a space $A$ directly.

	We now use $A$ to construct a cyclic covering.
	Let
	\[
	G=\ker\bigl(\Nrm_{E/K}:E^\times\to K^\times\bigr).
	\]
	The norm is onto, so $G$ has order
	$(Q^m-1)/(Q-1)=n$; it is cyclic as a subgroup of $E^\times$.
	Choose a generator $\beta$ of $G$.
	For every $y\in E^\times$, there is a unique
	$v\in A\setminus\{0\}$ with the same norm as $y$.
	Then $y/v\in G$, giving a unique expression
	$y=\beta^iv$ with $0\le i<n$.
	Consequently,
	\[
	E^\times=\bigsqcup_{i=0}^{n-1}
	\beta^i\bigl(A\setminus\{0\}\bigr).
	\]
	Thus the subspaces $\beta^iA$ cover $E$ and meet pairwise
	only in zero.
	
	To transfer this covering to $\F_q^n$, we use the powers of
	$\beta$. First note that $\F_q(\beta)=E$.
	Indeed,
	\[
	\ord(\beta)=n>q^{D-k}\ge q^{D/2},
	\]
	where the last inequality uses $m\ge2$.
	Every proper subfield of $E=\F_{q^D}$ containing $\F_q$
	has multiplicative group of order at most $q^{D/2}-1$,
	so none can contain $\beta$.
	Since $\beta^n=1$, the first $n$ powers of $\beta$ span $E$.
	Hence the linear map
	\begin{equation}\label{eq:explicit-perfect-covering}
		\pi:\F_q^n\longrightarrow E,\qquad
		\pi(z)=\sum_{i=0}^{n-1}z_i\beta^i
	\end{equation}
	is onto and satisfies $\pi\tau=\beta\pi$.
	
	Set $U=\pi^{-1}(A)$.
	The compatibility with the shift gives
	$\tau^iU=\pi^{-1}(\beta^iA)$.
	Pulling back the covering and intersection identities yields
	\[
	\begin{aligned}
		\bigcup_{i=0}^{n-1}\tau^iU
		&=\pi^{-1}\left(\bigcup_{i=0}^{n-1}\beta^iA\right)
		=\F_q^n,\\
		\core(U)
		&=\pi^{-1}\left(\bigcap_{i=0}^{n-1}\beta^iA\right)
		=\ker\pi.
	\end{aligned}
	\]
	Here the common intersection is zero because $n>1$ and the
	subspaces meet pairwise only in zero.
	Since $\pi$ is onto and $\dim_{\F_q}A=k$, we obtain
	\[
	\codim U=D-k=r,\qquad \codim\core(U)=D.
	\]
	
	Apply the normalized construction of
	Section~\ref{sec:structure} to $U$.
	The resulting cyclic code has syndrome dimension
	$D$ and alphabet dimension $D-r=k$.
	The identity
	\[
	n(q^k-1)=q^D-1
	\]
	gives equality in Lemma~\ref{lem:sphere-capacity}, so this
	code is perfect.
	Finally, the codimension-$r$ covering and the counting upper
	bound imply $h_q(n)=r$.
\end{proof}

The construction also gives an explicit parity-check matrix
for the resulting perfect code.
The map $\pi$ induces an isomorphism
$\F_q^n/\core(U)\cong E$, under which the quotient shift
corresponds to multiplication by $\beta$.
Fix an $\F_q$-basis of $E$, and let $M$ be the $D\times D$
matrix of this multiplication map.
Choose an $\F_q$-basis $b_1,\ldots,b_k$ of $A$, and let $B$
be the $D\times k$ matrix whose columns are their coordinates
in the chosen basis of $E$.

By the syndrome construction \eqref{eq:syndrome-construction},
the block corresponding to position $i$ is $M^iB$.
Thus the normalized code $P\le(\F_q^k)^n$ has parity-check matrix
\begin{equation}\label{eq:explicit-perfect-check}
	H=[\,B\ \ MB\ \ \cdots\ \ M^{n-1}B\,],\qquad
	P=\ker H,\qquad \rank H=D.
\end{equation}
Its dimension is therefore
$\dim_{\F_q}P=kn-D=k(n-m)$.

The prime-dimension case also yields further existence parameters
by restriction of scalars.

\begin{corollary}\label{cor:scalar-restriction}
Let $q$ be a prime power, let $k,m\ge2$, and put
$n=N_{q,k,m}$.
If there is a prime $p\mid k$ such that
$\gcd(m,q^{k/p}-1)=1$, then a cyclic $\F_q$-linear perfect
code of length $n$ over $\F_q^k$ exists, and
\[
h_q(n)=k(m-1).
\]
In particular, if $q$ is odd, $q+1$ is a power of two, and
$k=2p$ with $p$ prime, then for every $m\ge2$ such a code
exists if and only if $\gcd(m,q-1)=1$.
\end{corollary}

\begin{proof}
	Put $e=k/p$ and $q_0=q^e$.
	Apply the prime-dimension case of
	Theorem~\ref{thm:prime-block} over $\F_{q_0}$ with alphabet
	dimension $p$.
	Its proof supplies a $p$-dimensional $\F_{q_0}$-subspace
	$A$ of
	\[
	E=\F_{q_0^{pm}}=\F_{q^{km}}
	\]
	on which the norm to $K=\F_{q_0^p}=\F_{q^k}$ is bijective.
	Viewed over $\F_q$, the same space has dimension $ep=k$,
	and the norm is unchanged.
	The covering construction in that proof, now over $\F_q$,
	gives a cyclic perfect code and a cyclic covering of
	codimension $k(m-1)$.
	The counting bound then gives the stated value of $h_q(n)$.

	In the particular case, $q-1$ is even and $q+1$ has no
	prime divisor other than $2$. Therefore
	\[
	\gcd(m,q^2-1)=1\quad\Longleftrightarrow\quad
	\gcd(m,q-1)=1.
	\]
	Taking $e=2$ proves sufficiency.
	Necessity follows from the proof of
	Theorem~\ref{thm:prime-block}, which applies to every
	alphabet dimension.
\end{proof}

We conclude with the boundary case $m=2$, where attaining the
counting bound forces the normalized code to be perfect.
The following statement holds for every $k\ge1$.

\begin{corollary}\label{cor:boundary-application}\label{thm:boundary}
	Let $q$ be a prime power, let $k\ge1$, and put $n=q^k+1$.
	The following are equivalent:
	\begin{enumerate}[label=\textup{(\roman*)}]
		\item $h_q(n)=k$;
		\item there is a cyclic $\F_q$-linear perfect code in
		$(\F_q^k)^n$;
		\item $q$ is even.
	\end{enumerate}
	Moreover, every codimension-$k$ cyclic covering at this length
	has cyclic core of codimension $2k$, and its normalized syndrome
	code is perfect.
\end{corollary}

\begin{proof}
	Put $Q=q^k$, so that $n=Q+1$.
	The counting bound gives $h_q(n)\le k$.
	
	Let $U_0$ be any codimension-$k$ cyclic covering, and put
	$U_i=\tau^iU_0$ for $0\le i\le Q$.
	Each $U_i$ has dimension $n-k$, and the dimension formula gives
	\[
	\dim(U_i\cap U_0)\ge n-2k.
	\]
	Since the subspaces $U_i$ cover $\F_q^n$, counting their parts
	outside $U_0$ yields
	\[
	q^n
	\le |U_0|+\sum_{i=1}^{Q}|U_i\setminus U_0|
	\le q^{n-k}+Q\bigl(q^{n-k}-q^{n-2k}\bigr)
	=q^n.
	\]
	Both inequalities are therefore equalities.
	Thus the sets $U_i\setminus U_0$, $i>0$, are pairwise disjoint,
	and $\dim(U_i\cap U_0)=n-2k$ for every $i>0$.
	
	For distinct $i,j>0$, disjointness implies
	$U_i\cap U_j\subseteq U_i\cap U_0$.
	The left-hand side has dimension at least $n-2k$, while the
	right-hand side has exactly that dimension.
	Interchanging $i$ and $j$ consequently gives
	\[
	U_i\cap U_j=U_i\cap U_0=U_j\cap U_0.
	\]
	Since $Q\ge2$, all pairwise intersections are the same
	$(n-2k)$-dimensional space $C$, which is their common
	intersection $\core(U_0)$.
	
	The quotient $\F_q^n/C$ has dimension $2k$.
	Its subspaces $U_i/C$ have dimension $k$, cover the quotient,
	and meet pairwise only in zero.
	These are the one-block syndrome images of the normalized
	code, so Lemma~\ref{lem:perfect-syndromes} makes that code perfect.
	This proves \textup{(i)}$\Longrightarrow$\textup{(ii)} and
	the assertion about every codimension-$k$ covering.
	
	The necessity proof of Theorem~\ref{thm:prime-block} applies
	to every $k\ge1$.
	With $m=2$, it gives $\gcd(2,q-1)=1$, so $q$ is even.
	This proves \textup{(ii)}$\Longrightarrow$\textup{(iii)}.
	
	Finally, if $q$ is even, then $\gcd(2,q^k-1)=1$.
	Theorem~8 of~\cite{CER} therefore gives $h_q(q^k+1)=k$,
	proving \textup{(iii)}$\Longrightarrow$\textup{(i)}.
\end{proof}

\section{Conclusion}\label{sec:conclusion}

We characterized positive cyclic covering codimension by conditions
on cyclic $\F_q$-linear radius-one codes over vector alphabets.
The cyclic core supplies the syndrome quotient, and a one-generator
condition on the scalar dual gives the reverse construction.

For $k,m\ge2$ with $k$ prime or $m\ge2S_{q,k}+2$, we established
an arithmetic characterization: a cyclic $\F_q$-linear perfect
code of length $N_{q,k,m}$ over $\F_q^k$ exists if and only if
$\gcd(m,q-1)=1$.
When this condition holds, the norm construction yields cyclic
coverings attaining the counting bound
$h_q(N_{q,k,m})=k(m-1)$.
Restriction of scalars provides further existence results.
In particular, the same arithmetic criterion holds for every
$m\ge2$ when $q$ is odd, $q+1$ is a power of two, and
$k=2p$ with $p$ prime.
At the boundary length $q^k+1$, corresponding to $m=2$,
we proved for every $k\ge1$ that $h_q(q^k+1)=k$ if and only
if $q$ is even.
Every cyclic covering attaining this bound has cyclic core of
codimension $2k$ and produces a perfect normalized syndrome code.

These results lead to the following two questions.

\paragraph{Question 1.}
Let $q$ be a prime power and let $k,m\ge2$.
If $\gcd(m,q-1)=1$, does there always exist a cyclic
$\F_q$-linear perfect code of length
\[
N_{q,k,m}=\frac{q^{km}-1}{q^k-1}
\]
over $\F_q^k$?

\paragraph{Question 2.}
Let $q$ be a prime power and let $k,m\ge2$.
If $\gcd(m,q-1)>1$, is it always true that
\[
h_q(N_{q,k,m})\le k(m-1)-1?
\]

\section*{Funding}
This work was supported by the National Natural Science Foundation of China (grant numbers 12571003 and 12501006) and the Basic and Applied Basic Research Foundation of Guangdong Province (grant number 2024A1515010589).

\section*{AI Use Declaration}
During the preparation and revision of this manuscript, generative AI tools were used for literature searches, comparisons with related work, and finding examples. The mathematical claims, proofs, and final presentation remain the responsibility of the authors.

\section*{Data Availability}
No datasets were generated or analysed during the current study.

\section*{Competing Interests}
The authors declare no competing interests.

\setlength{\bibsep}{4pt plus 1pt minus 1pt}

\end{document}